\documentclass{llncs}

\usepackage{amsmath,amssymb,bm}
\usepackage{url}
\usepackage{isabelle,isabellesym}
\title{A Formalisation of Finite Automata using Hereditarily Finite Sets}
\author{Lawrence C. Paulson}
\institute{Computer Laboratory, University of Cambridge, England\\ \email{lp15@cam.ac.uk}}

\begin{document}
\maketitle

\begin{abstract}
Hereditarily finite (HF) set theory provides a standard universe of sets, but with no infinite sets.
Its utility is demonstrated through a formalisation of the theory of regular languages and finite automata, including the Myhill-Nerode theorem and Brzozowski's minimisation algorithm.
The states of an automaton are HF sets, possibly constructed by product, sum, powerset and similar operations.
\end{abstract}

\section{Introduction}

The theory of finite state machines is fundamental to computer science. It has applications to lexical analysis, hardware design and regular expression pattern matching. A regular language is one accepted by a finite state machine, or equivalently, one generated by a regular expression or a type-3 grammar \cite{Hopcroft-formal}.
Researchers have been formalising this theory for nearly three decades.

A critical question is how to represent the states of a machine.
Automata theory is developed using set-theoretic constructions, e.g.\ the product, disjoint sum or powerset of sets of states.
But in a strongly-typed formalism such as higher-order logic (HOL),  machines cannot be polymorphic in the type of states: statements such as ``every regular language is accepted by a finite state machine'' would require existential quantification over types.
One might conclude that there is no good way to formalise automata in HOL \cite{Doczkal-constructive,Wu-formalisation}. 

It turns out that finite automata theory can be formalised  within the theory of \emph{hereditarily finite sets}: set theory with the negation of the axiom of infinity.
It admits the usual constructions, including lists, functions and integers, but no infinite sets.
The type of HF sets can be constructed from the natural numbers within higher-order logic.
Using HF sets, we can retain the textbook definitions, without ugly numeric coding.
We can expect HF sets to find many other applications when formalising theoretical computer science.

The paper introduces HF set theory and automata (Sect.\ts\ref{sec:background}).
It presents a formalisation of deterministic finite automata and results such as the Myhill-Nerode theorem (Sect.\ts\ref{sec:dfa}).
It also treats nondeterministic finite automata and results such as the powerset construction and closure under  regular expression operations (Sect.\ts\ref{sec:nfa}).
Next come minimal automata, their uniqueness up to isomorphism, and Brzozowski's algorithm for minimising an automaton \cite{champarnaud-split} (Sect.\ts\ref{sec:minim}).
The paper concludes after discussing related work (Sect.\ts\ref{sec:related}--\ref{sec:concl}).
The proofs, which are available online \cite{Finite_Automata_HF-AFP}, also demonstrate the use of Isabelle's \emph{locales} \cite{ballarin-locales-module}.

\section{Background}
\label{sec:background}

An \emph{hereditarily finite set} can be understood inductively as a finite set of hereditarily finite sets  \cite{swierczkowski-finite}.
This definition justifies the recursive definition $f(x)=\sum\,\{2^{f(y)}\mid y\in x\}$, yielding a bijection $f:\text{HF}\to \mathbb{N}$  between the HF sets and the natural numbers.
The linear ordering on HF given by $x<y\iff f(x)<f(y)$ can be shown to extend both the membership and the subset relations.

The HF sets support many standard constructions, even quotients. Equivalence classes are not available in general --- they may be infinite --- but the linear ordering over HF identifies a unique representative.
The integers and rationals can be constructed, with their operations (but not the \emph{set} of integers, obviously).  
{\'S}wierczkowski \cite{swierczkowski-finite} has used HF as the basis for proving G\"odel's incompleteness theorems, and I have formalised his work using Isabelle \cite{paulson-incompl-ar}.

Let $\Sigma$ be a nonempty, finite alphabet of \emph{symbols}. Then $\Sigma^*$ is the set of \emph{words}: finite sequences of symbols. The empty word is written~$\epsilon$, and the concatenation of words $u$ and $v$ is written $uv$.
A \emph{deterministic finite automaton} (DFA) \cite{Hopcroft-formal,kozen-automata} is a structure $(K,\Sigma,\delta,q_0,F)$ where $K$ is a finite set of states, $\delta:K\times\Sigma\to K$ is the next-state function, $q_0\in K$ is the initial state and $F\subseteq K$ is the set of final or accepting states.
The next-state function on symbols is extended to one on words, $\delta^* : K\times\Sigma^*\to K$ such that $\delta^*(q,\epsilon)=q$,  $\delta^*(q,a)=\delta(q,a)$ for $a\in \Sigma$ and $\delta^*(q,uv)= \delta^*(\delta^*(q,u),v)$.
The DFA \emph{accepts} the string $w$ if $\delta^*(q_0,w)\in F$.
A set $L\subseteq\Sigma^*$ is a \emph{regular language} if $L$ is the set of strings accepted by some DFA\@.

A \emph{nondeterministic finite automaton} (NFA) is similar, but admits multiple execution paths and accepts a string if one of them reaches a final state. Formally, an NFA is a structure $(K,\Sigma,\delta,Q_0,F)$ where $\delta:K\times\Sigma\to \mathcal{P}(K)$ is the next-state function, $Q_0\subseteq K$ a set of initial states, the other components as above.
The next-state function is extended to $\delta^* : \mathcal{P}(K)\times\Sigma^*\to \mathcal{P}(K)$  such that $\delta^*(Q,\epsilon)=Q$,  $\delta^*(Q,a)=\bigcup_{q\in Q}\delta(q,a)$ for $a\in \Sigma$ and $\delta^*(Q,uv)= \delta^*(\delta^*(Q,u),v)$.
An NFA accepts the string $w$ provided $\delta^*(q,w)\in F$ for some $q\in Q_0$.

The notion of NFA can be extended with $\epsilon$-transitions, allowing ``silent'' transitions between states.
Define the transition relation $q\overset{a}\rightarrow q'$ for $q'\in \delta (q,a)$.
Let the $\epsilon$-transition relation  $q\overset{\epsilon}\rightarrow q'$ be given.
Then define the transition relation $q\overset{a}\Rightarrow q'$ to allow $\epsilon$-transitions before and after:
$(\overset{\epsilon}\rightarrow)^* \circ (\overset{a}\rightarrow) \circ (\overset{\epsilon}\rightarrow)^*$.

Every NFA can be transformed into a DFA, where the set of states is the powerset of the NFA's states, and the next-state function captures the effect of $q\overset{a}\Rightarrow q'$ on these sets of states.
Regular languages are closed under intersection and complement, therefore also under union. They are closed under repetition (Kleene star).
Two key results are discussed below:
\begin{itemize}
  \item The Myhill-Nerode theorem gives necessary and sufficient conditions for a language to be regular.
It defines a canonical and minimal DFA for any given regular language. Minimal DFAs are unique up to isomorphism.
  \item Reorienting the arrows of the transition relation transforms a DFA into an NFA accepting the reverse of the given language.
We can regain a DFA using the powerset construction.
Repeating this operation yields a minimal DFA for the original language.
This is Brzozowski's minimisation algorithm \cite{champarnaud-split}.
\end{itemize}

This work has been done using the proof assistant Isabelle/HOL\@.
Documentation is available online at \url{http://isabelle.in.tum.de/}.
The work refers to equivalence relations and equivalence classes, following the conventions established in my earlier paper \cite{paulson-equiv}.
If \isa{R} is an equivalence relation on the set \isa{A}, then \isa{A//R} is the set of equivalence classes. If \isa{x\isasymin A}, then its equivalence class is \isa{R``\{x\}}. 
Formally, it is the image of \isa{x} under \isa{R}: the set of all \isa{y} such that \isa{(x,y) \isasymin\ R}.
More generally, if \isa{X\isasymsubseteq A} then \isa{R``X} is the union of the equivalence classes \isa{R``\{x\}} for \isa{x\isasymin X}.

\section{Deterministic Automata; the Myhill-Nerode Theorem}
\label{sec:dfa}

When adopting HF set theory, there is the question of whether to use it for everything, or only where necessary.
The set of states is finite, so it could be an HF set, and similarly for the set of final states.
The alphabet could also be given by an HF set;
then words---lists of symbols---would also be HF sets.
Our definitions could be essentially typeless.

The approach adopted here is less radical. 
It makes a minimal use of HF, allowing stronger type-checking, although this does cause complications elsewhere.
Standard HOL sets (which are effectively predicates) are intermixed with HF sets.
An HF set has type \isa{hf}, while a (possibly infinite) set of HF sets has type \isa{hf~set}.
Definitions are polymorphic in the type \isa{'a} of alphabet symbols, while words have type \isa{'a~list}.

\subsection{Basic Definition of DFAs} \label{sec:DFAs}

The record definition below declares the components of a DFA\@. The types make it clear that there is indeed a set of states but only a single initial state, etc.

\begin{isabelle}
\isacommand{record}\ 'a\ dfa\ =\ states\ ::\ "hf\ set"\isanewline
\ \ \ \ \ \ \ \ \ \ \ \ \ \ \ \ init\ \ \ ::\ "hf"\isanewline
\ \ \ \ \ \ \ \ \ \ \ \ \ \ \ \ final\ \ ::\ "hf\ set"\isanewline
\ \ \ \ \ \ \ \ \ \ \ \ \ \ \ \ nxt\ \ \ \ ::\ "hf\ \isasymRightarrow \ 'a\ \isasymRightarrow \ hf"
\end{isabelle}

Now we package up the axioms of the DFA as a locale \cite{ballarin-locales-module}:
\begin{isabelle}
\isacommand{locale}\ dfa\ =\isanewline
\ \ \isakeyword{fixes}\ M\ ::\ "'a\ dfa"\isanewline
\ \ \isakeyword{assumes}\ init:\ \ "init\ M\ \isasymin \ states\ M"\isanewline
\ \ \ \ \ \ \isakeyword{and}\ final:\ \ "final\ M\ \isasymsubseteq \ states\ M"\isanewline
\ \ \ \ \ \ \isakeyword{and}\ nxt:\ \ \ \ "\isasymAnd q\ x.\ q\ \isasymin \ states\ M\ \isasymLongrightarrow \ nxt\ M\ q\ x\ \ \isasymin \ states\ M"\isanewline
\ \ \ \ \ \ \isakeyword{and}\ finite:\ "finite\ (states\ M)"
\end{isabelle}
The last assumption is needed because the \isa{states} field has type \isa{hf~set} and not~\isa{hf}.
The locale bundles the assumptions above into a local context, where they are directly available.
It is then easy to define the accepted language.
\begin{isabelle}
\isacommand{primrec}\ nextl\ ::\ "hf\ \isasymRightarrow\ 'a\ list\ \isasymRightarrow \ hf"\ \isakeyword{where}\isanewline
\ \ \ \ "nextl\ q\ []\ \ \ \ \ =\ q"\isanewline
\ \ |\ "nextl\ q\ (x\#xs)\ =\ nextl\ (nxt\ M\ q\ x)\ xs"
\vskip1mm
\isacommand{definition}\ language\ ::\ "('a\ list)\ set"\ \ \isakeyword{where}\isanewline
\ \ "language\ \isasymequiv \ \isacharbraceleft xs.\ nextl\ (init\ M)\ xs\ \isasymin \ final\ M\isacharbraceright "
\end{isabelle}
Equivalence relations play a significant role below. The following relation regards two
strings as equivalent if they take the machine to the same state \cite[p.\ts90]{kozen-automata}.
\begin{isabelle}
\isacommand{definition}\ eq\_nextl\ ::\ "('a\ list\ \isasymtimes \ 'a\ list)\ set"\ \isakeyword{where}\isanewline
\ \ "eq\_nextl\ \isasymequiv \ \isacharbraceleft (u,v).\ nextl\ (init\ M)\ u\ =\ nextl\ (init\ M)\ v\isacharbraceright "
\end{isabelle}
Note that \isa{language} and \isa{eq\_nextl} take no arguments, but refer to the locale.

\subsection{Myhill-Nerode Relations}

The Myhill-Nerode theorem asserts the equivalence of three characterisations of regular languages.
The first of these is to be the language accepted by some DFA\@. The other two are connected with certain equivalence relations, called Myhill-Nerode relations, on words of the language.

The definitions below are outside of the locale and are therefore independent of any particular DFA\@.
The predicate \isa{dfa} refers to the locale axioms and expresses that its argument, \isa{M}, is a DFA\@.
The predicate \isa{dfa.language} refers to the constant \isa{language}: outside of the locale, it takes a DFA as an argument.
\begin{isabelle}
\isacommand{definition}\ regular\ ::\ "('a\ list)\ set\ \isasymRightarrow \ bool"\ \isakeyword{where}\isanewline
\ \ "regular\ L\ \isasymequiv \ \isasymexists M.\ dfa\ M\ \isasymand \ dfa.language\ M\ =\ L"
\end{isabelle}
The other characterisations of a regular language involve abstract finite state machines derived from the language itself, with certain equivalence classes as the states.
A relation is \emph{right invariant} if it satisfies the following closure property.
\begin{isabelle}
\isacommand{definition}\ right\_invariant\ ::\ "('a\ list\ \isasymtimes \ 'a\ list)\ set\ \isasymRightarrow \ bool"\ \isakeyword{where}\isanewline
\ \ "right\_invariant\ r\ \isasymequiv \ (\isasymforall u\ v\ w.\ (u,v)\ \isasymin \ r\ \isasymlongrightarrow \ (u@w,\ v@w)\ \isasymin \ r)"
\end{isabelle}
The intuition is that if two words \isa{u} and \isa{v} are related, then each word brings the ``machine'' to the same state, and once this has happened, this agreement must continue no matter how the words are extended as \isa{u@w} and \isa{v@w}.

A \emph{Myhill-Nerode relation} for a language \isa{L} is a right invariant equivalence relation of finite index where \isa{L} is the union of some of the equivalence classes \cite[p.\ts90]{kozen-automata}.
\emph{Finite index} means the set of equivalence classes is finite: \isa{finite\ (UNIV//R)}.%
\footnote{\isa{UNIV} denotes a typed universal set, here the set of all words.}
The equivalence classes will be the states of a finite state machine.
The equality \isa{L\ =\ R``A}, where \isa{A\ \isasymsubseteq\ L} is a set of words of the language, expresses \isa{L} as the union of a set of equivalence classes, which will be the final states.

\begin{isabelle}
\isacommand{definition}\ MyhillNerode\ ::\ "'a\ list\ set\ \isasymRightarrow \ ('a\ list\ *\ 'a\ list)set\ \isasymRightarrow \ bool"\isanewline
\ \ \isakeyword{where}\ \ "MyhillNerode\ L\ R\ \isasymequiv \ equiv\ UNIV\ R\ \isasymand \ right\_invariant\ R\ \isasymand\isanewline
\ \ \ \ \ \ \ \ \ \ \ \ \ \ \ \ \ \ \ \ \ \ \ \ \ \ \ \ \ \ \ finite\ (UNIV//R)\ \isasymand \ (\isasymexists A.\ L\ =\ R``A)"
\end{isabelle}
While \isa{eq\_nextl} (defined in \S\ref{sec:DFAs}) refers to a machine, the relation \isa{eq\_app\_right} is defined in terms of a language, \isa{L}\@. It relates the words \isa{u} and \isa{v} if all extensions of them, \isa{u@w} and \isa{v@w}, behave equally with respect to \isa{L}\@:
\begin{isabelle}
\isacommand{definition}\ eq\_app\_right\ ::\ "'a\ list\ set\ \isasymRightarrow \ ('a\ list\ *\ 'a\ list)\ set"\ \isakeyword{where}\isanewline
\ \ "eq\_app\_right\ L\ \isasymequiv \ \isacharbraceleft (u,v).\ \isasymforall w.\ u@w\ \isasymin \ L\ \isasymlongleftrightarrow \ v@w\ \isasymin \ L\isacharbraceright "
\end{isabelle}
It is a Myhill-Nerode relation for \isa{L} provided it is of finite index:
\begin{isabelle}
\isacommand{lemma}\ MN\_eq\_app\_right:\isanewline
\ \ \ \ \ "finite\ (UNIV\ //\ eq\_app\_right\ L)\ \isasymLongrightarrow \ MyhillNerode\ L\ (eq\_app\_right\ L)"
\end{isabelle}
Moreover, every Myhill-Nerode relation \isa{R} for \isa{L} refines \isa{eq\_app\_right\ L}. 
\begin{isabelle}
\isacommand{lemma}\ MN\_refines\_eq\_app\_right:\ "MyhillNerode\ L\ R\ \isasymLongrightarrow \ R\ \isasymsubseteq \ eq\_app\_right\ L"
\end{isabelle}
This essentially states that \isa{eq\_app\_right~L} is the most abstract Myhill-Nerode relation for~\isa{L}\@.
This will eventually yield a way of defining a minimal machine.

\subsection{The Myhill-Nerode Theorem}

The Myhill-Nerode theorem says that these three statements are equivalent \cite{Hopcroft-formal}:
\begin{enumerate}
  \item The set \isa{L} is a regular language (is accepted by some DFA).
  \item There exists some Myhill-Nerode relation \isa{R} for \isa{L}.
  \item The relation \isa{eq\_app\_right L} has finite index.
\end{enumerate}
We have $(1)\Rightarrow(2)$ because \isa{eq\_nextl} is a Myhill-Nerode relation.
We have $(2)\Rightarrow(3)$, by lemma \isa{MN\_refines\_eq\_app\_right}, because every equivalence class for \isa{eq\_app\_right L} is the union of equivalence classes of \isa{R}, and so \isa{eq\_app\_right\ L} has minimal index for all Myhill-Nerode relations.
We get $(3)\Rightarrow(1)$ by constructing a DFA whose states are the (finitely many) equivalence classes of \isa{eq\_app\_right L}. This construction can be done for every Myhill-Nerode relation.

Until now, all proofs have been routine. But now we face a difficulty: the states of our machine should be equivalence classes of words, but these could be infinite sets.
What can be done?
The solution adopted here is to map the equivalence classes to the natural numbers, which are easily embedded in HF\@.
Proving that the set of equivalence classes is finite gives us such a map.

Mapping infinite sets to integers seems to call into question the very idea of representing states by HF sets.
However, mapping sets to integers turns out to be convenient only occasionally, and it is not necessary:
we could formalise DFAs differently, coding symbols (and therefore words) as HF sets.
Then we could represent states by representatives (having type \isa{hf}) of equivalence classes.
Using Isabelle's type-class system to identify the types (integers, booleans, lists, etc.) that can be embedded into HF,
type \isa{'a\ dfa} could still be polymorphic in the type of symbols.
But the approach followed here is simpler.

\subsection{Constructing a DFA from a Myhill-Nerode Relation}
\label{sec:canon_dfa}

If \isa{R} is a Myhill-Nerode relation for a language \isa{L}, then the set of equivalence classes is finite and yields a DFA for~\isa{L}\@.
The construction is packaged as a locale, which is used once in the proof of the Myhill-Nerode theorem, and again to prove that minimal DFAs are unique.
The locale includes not only \isa{L} and \isa{R}, but also the set \isa{A} of accepting states, the cardinality \isa{n} and the bijection \isa{h} between the set \isa{UNIV//R} of equivalence classes and the number \isa{n} as represented in HF\@.
The locale assumes the Myhill-Nerode conditions.
\begin{isabelle}
\isacommand{locale}\ MyhillNerode\_dfa\ =\isanewline
\ \ \isakeyword{fixes}\ L\ ::\ "('a\ list)\ set"\ \isakeyword{and}\ R\ ::\ "('a\ list\ *\ 'a\ list)\ set"\ \isanewline
\ \ \ \isakeyword{and}\ A\ ::\ "('a\ list)\ set"\ \isakeyword{and}\ n\ ::\ nat\ \isakeyword{and}\ h\ ::\ "('a\ list)\ set\ \isasymRightarrow \ hf"\isanewline
\ \ \isakeyword{assumes}\ eqR:\ "equiv\ UNIV\ R"\isanewline
\ \ \ \ \ \ \isakeyword{and}\ riR:\ "right\_invariant\ R"\isanewline
\ \ \ \ \ \ \isakeyword{and}\ L:\ \ \ "L\ =\ R``A"\isanewline
\ \ \ \ \ \ \isakeyword{and}\ h:\ \ \ "bij\_betw\ h\ (UNIV//R)\ (hfset\ (ord\_of\ n))"
\end{isabelle}
The DFA is defined within the locale.
The states are given by the equivalence classes.
The initial state is the equivalence class for the empty word; the set of final states is derived from the set \isa{A} of words that generate \isa{L};  the next-state function maps the equivalence class for the word \isa{u} to that for \isa{u@[x]}.
Equivalence classes are not the actual states here, but are mapped to integers via the bijection~\isa{h}.
As mentioned above, this use of integers is not essential.
\begin{isabelle}
\ \ \isacommand{definition}\ DFA\ ::\ "'a\ dfa"\ \isakeyword{where}\isanewline
\ \ \ \ "DFA\ =\ \isasymlparr states\ =\ h\ `\ (UNIV//R),\isanewline
\ \ \ \ \ \ \ \ \ \ \ \ init\ \ \ =\ h\ (R\ ``\ \isacharbraceleft []\isacharbraceright ),\isanewline
\ \ \ \ \ \ \ \ \ \ \ \ final\ \ =\ \isacharbraceleft h\ (R\ ``\ \isacharbraceleft u\isacharbraceright )\ |\ u.\ u\ \isasymin \ A\isacharbraceright ,\isanewline
\ \ \ \ \ \ \ \ \ \ \ \ nxt\ \ \ \ =\ \isasymlambda q\ x.\ h\ (\isasymUnion u\ \isasymin \ h\isactrlsup -\isactrlsup 1\ q.\ R\ ``\ \isacharbraceleft u@[x]\isacharbraceright )\isasymrparr "
\end{isabelle}
This can be proved to be a DFA easily.
One proof line, using the right-invariance property and lemmas about quotients \cite{paulson-equiv}, proves that the next-state function respects the equivalence relation.
Four more lines are needed to verify the properties of a DFA, somewhat more to show that the language of this DFA is indeed~\isa{L}\@.

The facts proved within the locale are summarised (outside its scope) by the following theorem, stating that every Myhill-Nerode relation yields an equivalent DFA\@. (The \isakeyword{obtains} form expresses existential and multiple conclusions.)
\begin{isabelle}
\isacommand{theorem}\ MN\_imp\_dfa:\isanewline
\ \ \isakeyword{assumes}\ "MyhillNerode\ L\ R"\isanewline
\ \ \isakeyword{obtains}\ M\ \isakeyword{where}\ "dfa\ M"\ \ "dfa.language\ M\ =\ L"\isanewline
\ \ \ \ \ \ \ \ \ \ \ \ \ \ \ \ \ \ \ "card\ (states\ M)\ =\ card\ (UNIV//R)"
\end{isabelle}
This completes the $(3)\Rightarrow(1)$ stage, by far the hardest, of the  Myhill-Nerode theorem.
The three stages are shown below. 
Lemma \isa{L2\_3} includes a result about cardinality: the construction yields a minimal DFA, which will be useful later.
\begin{isabelle}
\isacommand{lemma}\ L1\_2:\ "regular\ L\ \isasymLongrightarrow \ \isasymexists R.\ MyhillNerode\ L\ R"
\vskip1mm
\isacommand{lemma}\ L2\_3:\isanewline
\ \ \isakeyword{assumes}\ "MyhillNerode\ L\ R"\isanewline
\ \ \isakeyword{obtains}\ "finite\ (UNIV\ //\ eq\_app\_right\ L)"\isanewline
\ \ \ \ \ \ \ \ \ \ "card\ (UNIV\ //\ eq\_app\_right\ L)\ \isasymle \ card\ (UNIV\ //\ R)"
\vskip1mm
\isacommand{lemma}\ L3\_1:\ "finite\ (UNIV\ //\ eq\_app\_right\ L)\ \isasymLongrightarrow \ regular\ L"
\end{isabelle}

\section{Nondeterministic Automata and Closure Proofs}
\label{sec:nfa}

As most of the proofs are simple, our focus will be the use of HF sets when defining automata.
Our main example is the powerset construction for transforming a nondeterministic automaton into a deterministic one.

\subsection{Basic Definition of NFAs} \label{sec:NFAs}

As in the deterministic case, a record holds the necessary components, while a locale encapsulates the axioms.
Component \isa{eps} deals with $\epsilon$-transitions. 
\begin{isabelle}
\isacommand{record}\ 'a\ nfa\ =\ states\ ::\ "hf\ set"\isanewline
\ \ \ \ \ \ \ \ \ \ \ \ \ \ \ \ init\ \ \ ::\ "hf\ set"\isanewline
\ \ \ \ \ \ \ \ \ \ \ \ \ \ \ \ final\ \ ::\ "hf\ set"\isanewline
\ \ \ \ \ \ \ \ \ \ \ \ \ \ \ \ nxt\ \ \ \ ::\ "hf\ \isasymRightarrow \ 'a\ \isasymRightarrow \ hf\ set"\isanewline
\ \ \ \ \ \ \ \ \ \ \ \ \ \ \ \ eps\ \ \ \ ::\ "(hf\ *\ hf)\ set"
\end{isabelle}
The axioms are obvious: the initial, final and next states belong to the set of states, which is finite.
An axiom restricting $\epsilon$-transitions to machine states was removed, as it did not simplify proofs.
Working with $\epsilon$-transitions is messy.
It helps to provide special treatment for NFAs having no $\epsilon$-transitions.
Allowing multiple initial states reduces the need for $\epsilon$-transitions.
\begin{isabelle}
\isacommand{locale}\ nfa\ =\isanewline
\ \ \isakeyword{fixes}\ M\ ::\ "'a\ nfa"\isanewline
\ \ \isakeyword{assumes}\ init:\ "init\ M\ \isasymsubseteq \ states\ M"\isanewline
\ \ \ \ \ \ \isakeyword{and}\ final:\ "final\ M\ \isasymsubseteq \ states\ M"\isanewline
\ \ \ \ \ \ \isakeyword{and}\ nxt:\ \ \ "\isasymAnd q\ x.\ q\ \isasymin \ states\ M\ \isasymLongrightarrow \ nxt\ M\ q\ x\ \isasymsubseteq \ states\ M"\isanewline
\ \ \ \ \ \ \isakeyword{and}\ finite:\ "finite\ (states\ M)"
\end{isabelle}
The following function ``closes up'' a set \isa{Q} of states under $\epsilon$-transitions.
Intersection with \isa{states\ M} confines these transitions to legal states.
\begin{isabelle}
\isacommand{definition}\ epsclo\ ::\ "hf\ set\ \isasymRightarrow \ hf\ set"\ \isakeyword{where}\isanewline
\ \ "epsclo\ Q\ \isasymequiv \ states\ M\ \isasyminter \ (\isasymUnion q\isasymin Q.\ \isacharbraceleft q'.\ (q,q')\ \isasymin \ (eps\ M)\isactrlsup *\isacharbraceright )"
\end{isabelle}
The remaining definitions are straightforward. Note that \isa{nextl} generalises \isa{nxt} to take a set of states as well is a list of symbols.
\begin{isabelle}
\isacommand{primrec}\ nextl\ ::\ "hf\ set\ \isasymRightarrow\ 'a\ list\ \isasymRightarrow \ hf\ set"\ \isakeyword{where}\isanewline
\ \ \ \ "nextl\ Q\ []\ \ \ \ \ =\ epsclo\ Q"\isanewline
\ \ |\ "nextl\ Q\ (x\#xs)\ =\ nextl\ (\isasymUnion q\ \isasymin \ epsclo\ Q.\ nxt\ M\ q\ x)\ xs"
\vskip1mm
\isacommand{definition}\ language\ ::\ "('a\ list)\ set"\ \ \isakeyword{where}\isanewline
\ \ "language\ \isasymequiv \ \isacharbraceleft xs.\ nextl\ (init\ M)\ xs\ \isasyminter \ final\ M\ \isasymnoteq \ \isacharbraceleft \isacharbraceright \isacharbraceright "
\end{isabelle}

\subsection{The Powerset Construction}

The construction of a DFA to simulate a given NFA is elementary, and is a good demonstration of the HF sets.
The strongly-typed approach used here requires a pair of coercion functions \isa{hfset\ ::\ "hf\ \isasymRightarrow \ hf\ set"} and \isa{HF\ ::\ "hf\ set\ \isasymRightarrow \ hf"} to convert between HF sets and ordinary sets.
\begin{isabelle}
\isacommand{lemma}\ HF\_hfset:\ "HF\ (hfset\ a)\ =\ a"\isanewline
\isacommand{lemma}\ hfset\_HF:\ "finite\ A\ \isasymLongrightarrow \ hfset\ (HF\ A)\ =\ A"
\end{isabelle}
With this approach, type-checking indicates whether we are dealing with a set of states or a single state.
The drawback is that we occasionally have to show that a set of states is finite in the course of reasoning about the coercions, which would never be necessary if we confined our reasoning to the HF world.

Here is the definition of the DFA\@. 
The states are $\epsilon$-closed subsets of NFA states, coerced to type \isa{hf}.
The initial and final states are defined similarly, while the next-state function requires both coercions and performs $\epsilon$-closure before and after. We work in locale \isa{nfa}, with access to the components of the NFA\@.
\begin{isabelle}
\isacommand{definition}\ Power\_dfa\ ::\ "'a\ dfa"\ \isakeyword{where}\isanewline
\ "Power\_dfa\ =\ \isasymlparr dfa.states\ =\ HF\ `\ epsclo\ `\ Pow\ (states\ M),\isanewline
\ \ \ \ \ \ \ \ \ \ \ init\ \ =\ HF(epsclo(init\ M)),\isanewline
\ \ \ \ \ \ \ \ \ \ \ final\ =\ \isacharbraceleft HF(epsclo\ Q)\ |\ Q.\ Q\ \isasymsubseteq \ states\ M\ \isasymand \ Q\ \isasyminter \ final\ M\ \isasymnoteq \ \isacharbraceleft \isacharbraceright \isacharbraceright ,\isanewline
\ \ \ \ \ \ \ \ \ \ \ nxt\ \ \ =\ \isasymlambda Q\ x.\ HF(\isasymUnion q\ \isasymin \ epsclo\ (hfset\ Q).\ epsclo\ (nxt\ M\ q\ x))\isasymrparr "
\end{isabelle}
Proving that this is a DFA is trivial. The hardest case is to show that the next-state function maps states to states.
%
%
Proving that the two automata accept the same language is also simple, by reverse induction on lists (the induction step concerns \isa{u@[x]}, putting \isa{x} at the end). Here, \isa{Power.language} refers to the language of the powerset DFA, while \isa{language} refers to that of the NFA\@.

\begin{isabelle}
\isacommand{theorem}\ Power\_language:\ "Power.language\ =\ language"
\end{isabelle}

\subsection{Other Closure Properties}

The set of languages accepted by some DFA is closed under complement, intersection, concatenation, repetition (Kleene star), etc. \cite{Hopcroft-formal}.
Consider intersection:
\begin{isabelle}
\isacommand{theorem}\ regular\_Int:\isanewline
\ \ \isakeyword{assumes}\ S:\ "regular\ S"\ \isakeyword{and}\ T:\ "regular\ T"\ \isakeyword{shows}\ "regular\ (S\ \isasyminter \ T)"
\end{isabelle}
The recognising DFA is created by forming the Cartesian product of the sets of states of \isa{MS} and \isa{MT}, the DFAs of the two languages. The machines are effectively run in parallel.
The decision to represent a set of states by type \isa{hf~set} rather than by type~\isa{hf} means we cannot write \isa{dfa.states\ MS\ \isasymtimes \ dfa.states\ MT}, but we can express this concept using set comprehension:
\begin{isabelle}
\ \ "\isasymlparr states\ =\ \isacharbraceleft \isasymlangle q1,q2\isasymrangle \ |\ q1\ q2.\ q1\ \isasymin \ dfa.states\ MS\ \isasymand \ q2\ \isasymin \ dfa.states\ MT\isacharbraceright ,\isanewline
\ \ \ \ init\ \ \ =\ \isasymlangle dfa.init\ MS,\ dfa.init\ MT\isasymrangle ,\isanewline
\ \ \ \ final\ \ =\ \isacharbraceleft \isasymlangle q1,q2\isasymrangle \ |\ q1\ q2.\ q1\ \isasymin \ dfa.final\ MS\ \isasymand \ q2\ \isasymin \ dfa.final\ MT\isacharbraceright ,\isanewline
\ \ \ \ nxt\ \ \ \ =\ \isasymlambda \isasymlangle qs,qt\isasymrangle \ x.\ \isasymlangle dfa.nxt\ MS\ qs\ x,\ dfa.nxt\ MT\ qt\ x\isasymrangle \isasymrparr "
\end{isabelle}
This is trivially shown to be a DFA\@. Showing that it accepts the intersection of the given languages is again easy by reverse induction.

Closure under concatenation is expressed as follows:
\begin{isabelle}
\isacommand{theorem}\ regular\_conc:\isanewline
\ \ \isakeyword{assumes}\ S:\ "regular\ S"\ \isakeyword{and}\ T:\ "regular\ T"\ \isakeyword{shows}\ "regular\ (S\ @@\ T)"
\end{isabelle}

The concatenation is recognised by an NFA involving the disjoint sum of the sets of states of \isa{MS} and \isa{MT}, the DFAs of the two languages. The effect is to simulate the first machine until it accepts a string, then to transition to a simulation of the second machine.
There are $\epsilon$-transitions linking every final state of~\isa{MS} to the initial state of~\isa{MT}\@.
We again cannot write \isa{dfa.states\ MS\ + \ dfa.states\ MT}, but we can express the disjoint sum naturally enough:
\begin{isabelle}
\ \ "\isasymlparr states\ =\ Inl\ `\ (dfa.states\ MS)\ \isasymunion \ Inr\ `\ (dfa.states\ MT),\isanewline
\ \ \ \ \ init\ \ =\ \isacharbraceleft Inl\ (dfa.init\ MS)\isacharbraceright ,\isanewline
\ \ \ \ \ final\ =\ Inr\ `\ (dfa.final\ MT),\isanewline
\ \ \ \ \ nxt\ \ \ =\ \isasymlambda q\ x.\ sum\_case\ (\isasymlambda qs.\ \isacharbraceleft Inl\ (dfa.nxt\ MS\ qs\ x)\isacharbraceright )\isanewline
\ \ \ \ \ \ \ \ \ \ \ \ \ \ \ \ \ \ \ \ \ \ \ \ \ \ \ \ (\isasymlambda qt.\ \isacharbraceleft Inr\ (dfa.nxt\ MT\ qt\ x)\isacharbraceright )\ q,\isanewline
\ \ \ \ \ eps\ \ \ =\ (\isasymlambda q.\ (Inl\ q,\ Inr\ (dfa.init\ MT)))\ `\ dfa.final\ MS\isasymrparr "
\end{isabelle}
Again, it is trivial to show that this is an NFA\@. But unusually, proving that it recognises the concatenation of the languages is a challenge.
We need to show, by induction, that the ``left part'' of the NFA correctly simulates \isa{MS}\@.
\begin{isabelle}
\ \ \ \ \isacommand{have}\ "\isasymAnd q.\ Inl\ q\ \isasymin \ ST.nextl\ \isacharbraceleft Inl\ (dfa.init\ MS)\isacharbraceright \ u\ \isasymlongleftrightarrow\isanewline
\ \ \ \ \ \ \ \ \ \ \ \ \ \ \ q\ =\ (dfa.nextl\ MS\ (dfa.init\ MS)\ u)"
\end{isabelle}
The key property is that any string accepted by the NFA can be split into strings accepted by the two DFAs.
The proof involves a fairly messy induction.
\begin{isabelle}
\ \ \ \ \isacommand{have}\ "\isasymAnd q.\ Inr\ q\ \isasymin \ ST.nextl\ \isacharbraceleft Inl\ (dfa.init\ MS)\isacharbraceright \ u \isasymlongleftrightarrow\isanewline
\ \ \ \ \ \ \ \ \ \ \ \ \ \ \ (\isasymexists uS\ uT.\ uS\ \isasymin \ dfa.language\ MS\ \isasymand \ u\ =\ uS@uT\ \isasymand \ \isanewline
\ \ \ \ \ \ \ \ \ \ \ \ \ \ \ \ \ \ \ \ \ \ \ \ \ q\ =\ dfa.nextl\ MT\ (dfa.init\ MT)\ uT)"
\end{isabelle}

%
%
Closure under Kleene star is not presented here, as it involves no interesting set operations. 
The language $L^*$ is recognised by an NFA with an extra state, which serves as the initial state and runs the DFA for~$L$ including iteration.
The proofs are messy, with many cases.
To their credit, Hopcroft and Ullman \cite{Hopcroft-formal} give some details, while other authors content themselves with diagrams alone.

\section{State Minimisation for DFAs}
\label{sec:minim}

Given a regular language~$L$, the Myhill-Nerode theorem yields a DFA having the minimum number of states.
But it does not yield a minimisation algorithm for a given automaton.
It turns out that a DFA is minimal if it has no unreachable states and if no two states are \emph{indistinguishable} (in a sense made precise below).
This again does not yield an algorithm.
\emph{Brzozowski's minimisation algorithm} involves reversing the DFA to create an NFA, converting back to a DFA via powersets, removing unreachable states, then repeating those steps to undo the reversal.
Surprisingly, it performs well in practice \cite{champarnaud-split}.

\subsection{The Left and Right Languages of a State}

The following developments are done within the locale \isa{dfa}, and therefore refer to one particular deterministic finite automaton.

The \emph{left language} of a state~$q$ is the set of all words $w$ such that $q_0 \overset{w}{\rightarrow^*} q$, or informally, such that the machine when started in the initial state and given the word~$w$ ends up in~$q$.
In a DFA, the left languages of distinct states are disjoint, if they are nonempty.
\begin{isabelle}
\isacommand{definition}\ left\_lang\ ::\ "hf\ \isasymRightarrow \ ('a\ list)\ set"\ \ \isakeyword{where}\isanewline
\ \ "left\_lang\ q\ \isasymequiv \ \isacharbraceleft u.\ nextl\ (init\ M)\ u\ =\ q\isacharbraceright "
\end{isabelle}

The \emph{right language} of a state~$q$ is the set of all words $w$ such that $q \overset{w}{\rightarrow^*} q_f$, where $q_f$ is a final state, or informally, such that the machine when started in $q$ will accept the word~$w$.
The language of a DFA is the right language of~$q_0$.
Two states having the same right language are \emph{indistinguishable}: they both lead to the same  words being accepted.
\begin{isabelle}
\isacommand{definition}\ right\_lang\ ::\ "hf\ \isasymRightarrow \ ('a\ list)\ set"\ \ \isakeyword{where}\isanewline
\ \ "right\_lang\ q\ \isasymequiv \ \isacharbraceleft u.\ nextl\ q\ u\ \isasymin \ final\ M\isacharbraceright"
\end{isabelle}

The \emph{accessible} states are those that can be reached by at least one word.
\begin{isabelle}
\isacommand{definition}\ accessible\ ::\ "hf\ set"\ \ \isakeyword{where}\isanewline
\ \ "accessible\ \isasymequiv \ \isacharbraceleft q.\ left\_lang\ q\ \isasymnoteq \ \isacharbraceleft \isacharbraceright \isacharbraceright "
\end{isabelle}

The function \isa{path\_to} returns one specific such word.
This function will eventually be used to express an isomorphism between any minimal DFA (one having no inaccessible or indistinguishable states) and the canonical DFA determined by the Myhill-Nerode theorem.
\begin{isabelle}
\isacommand{definition}\ path\_to\ ::\ "hf\ \isasymRightarrow \ 'a\ list"\ \ \isakeyword{where}\isanewline
\ \ "path\_to\ q\ \isasymequiv \ SOME\ u.\ u\ \isasymin \ left\_lang\ q"
\vskip1mm
\isacommand{lemma}\ nextl\_path\_to:\isanewline
\ \ "q\ \isasymin \ accessible\ \isasymLongrightarrow \ nextl\ (dfa.init\ M)\ (path\_to\ q)\ =\ q"
\end{isabelle}

First, we deal with the problem of inaccessible states. It is easy to restrict any DFA to one having only accessible states.
\begin{isabelle}
\isacommand{definition}\ Accessible\_dfa\ ::\ "'a\ dfa"\ \isakeyword{where}\isanewline
\ \ "Accessible\_dfa\ =\ \isasymlparr dfa.states\ =\ accessible,\isanewline
\ \ \ \ \ \ \ \ \ \ \ \ \ \ \ \ \ \ \ \ \ init\ \ =\ init\ M,\isanewline
\ \ \ \ \ \ \ \ \ \ \ \ \ \ \ \ \ \ \ \ \ final\ =\ final\ M\ \isasyminter \ accessible,\isanewline
\ \ \ \ \ \ \ \ \ \ \ \ \ \ \ \ \ \ \ \ \ nxt\ \ \ =\ nxt\ M\isasymrparr "
\end{isabelle}

This construction is readily shown to be a DFA that agrees with the original in most respects.
In particular, the two automata agree on \isa{left\_lang} and \isa{right\_lang}, and therefore on the language they accept:
\begin{isabelle}
\isacommand{lemma}\ Accessible\_language:\ "Accessible.language\ =\ language"
\end{isabelle}
We can now define a DFA to be minimal if all states are accessible and no two states have the same right language.
(The formula \isa{inj\_on\ right\_lang\ (dfa.states\ M)} expresses that the function \isa{right\_lang} is injective on the set \isa{dfa.states\ M}.)
\begin{isabelle}
\isacommand{definition}\ minimal\ \isakeyword{where}\isanewline
\ \ "minimal\ \isasymequiv \ accessible\ =\ states\ M\ \isasymand \ inj\_on\ right\_lang\ (dfa.states\ M)"
\end{isabelle}
Because we are working within the DFA locale, \isa{minimal} is a constant referring to one particular automaton.

\subsection{￼A Collapsing Construction}

We can deal with indistinguishable states similarly, defining a DFA in which the indistinguishable states are identified via equivalence classes.
This is not part of Brzozowski's minimisation algorithm, but it is interesting in its own right: the equivalence classes themselves are HF sets.
We begin by declaring a relation stating that two states are equivalent if they have the same right language.
\begin{isabelle}
\isacommand{definition}\ eq\_right\_lang\ ::\ "(hf\ \isasymtimes \ hf)\ set"\ \isakeyword{where}\isanewline
\ \ "eq\_right\_lang\ \isasymequiv \ \isacharbraceleft (u,v).\ u\ \isasymin \ states\ M\ \isasymand \ v\ \isasymin \ states\ M\ \isasymand\isanewline
\ \ \ \ \ \ \ \ \ \ \ \ \ \ \ \ \ \ \ \ \ \ \ \ \ \ \ right\_lang\ u\ =\ right\_lang\ v\isacharbraceright "
\end{isabelle}

Trivially, this is an equivalence relation, and equivalence classes of states are finite (there are only finitely many states).
In the corresponding DFA, these equivalence classes form the states, with the initial and final states given by the equivalence classes for the corresponding states of the original DFA\@.
As usual, the function~\isa{HF} is used to coerce a set of states to type~\isa{hf}.
\begin{isabelle}
\isacommand{definition}\ Collapse\_dfa\ ::\ "'a\ dfa"\ \isakeyword{where}\isanewline
\ \ "Collapse\_dfa\ =\ \isasymlparr dfa.states\ =\ HF\ `\ (states\ M\ //\ eq\_right\_lang),\isanewline
\ \ \ \ \ \ \ \ \ \ \ \ \ \ \ \ \ \ \ init\ \ =\ HF\ (eq\_right\_lang\ ``\ \isacharbraceleft init\ M\isacharbraceright ),\isanewline
\ \ \ \ \ \ \ \ \ \ \ \ \ \ \ \ \ \ \ final\ =\ \isacharbraceleft HF\ (eq\_right\_lang\ ``\ \isacharbraceleft q\isacharbraceright )\ |\ q.\ q\ \isasymin \ final\ M\isacharbraceright ,\isanewline
\ \ \ \ \ \ \ \ \ \ \ \ nxt\ =\ \isasymlambda Q\ x.\ HF\ (\isasymUnion q\ \isasymin \ hfset\ Q.\ eq\_right\_lang\ ``\ \isacharbraceleft nxt\ M\ q\ x\isacharbraceright )\isasymrparr "
\end{isabelle}
This is easily shown to be a DFA, and the next-state function respects the equivalence relation.
Showing that it accepts the same language is straightforward.
%
%
\begin{isabelle}
\isacommand{lemma}\ ext\_language\_Collapse\_dfa:\isanewline
\ \ \ \ \ "u\ \isasymin \ Collapse.language\ \isasymlongleftrightarrow \ u\ \isasymin \ language"
\end{isabelle}

\subsection{The Uniqueness of Minimal DFAs}

The property \isa{minimal} is true for machines having no inaccessible or indistinguishable states.
To prove that such a machine actually has a minimal number of states is tricky. It can be shown to be isomorphic to the canonical machine from the Myhill-Nerode theorem, which indeed has a minimal number of states.

Automata \isa{M} and \isa{N} are \emph{isomorphic} if there exists a bijection \isa{h} between their state sets that preserves their initial, final and next states.
This conception is nicely captured by a locale, taking the DFAs as parameters: 
\begin{isabelle}
\isacommand{locale}\ dfa\_isomorphism\ =\ M:\ dfa\ M\ +\ N:\ dfa\ N\isanewline
\ \ \ \ \ \ \ \ \ \ \ \ \ \ \ \ \ \ \ \ \ \ \ \ \isakeyword{for}\ M\ ::\ "'a\ dfa"\ \isakeyword{and}\ N\ ::\ "'a\ dfa"\ +\isanewline
\ \ \isakeyword{fixes}\ h\ ::\ "hf\ \isasymRightarrow \ hf"\isanewline
\ \ \isakeyword{assumes}\ h:\ "bij\_betw\ h\ (states\ M)\ (states\ N)"\isanewline
\ \ \ \ \ \ \isakeyword{and}\ init\ :\ "h\ (init\ M)\ =\ init\ N"\isanewline
\ \ \ \ \ \ \isakeyword{and}\ final:\ "h\ `\ final\ M\ =\ final\ N"\isanewline
\ \ \ \ \ \ \isakeyword{and}\ nxt\ \ :\ "\isasymAnd q\ x.\ q\ \isasymin \ states\ M\ \isasymLongrightarrow \ h(nxt\ M\ q\ x)\ =\ nxt\ N\ (h\ q)\ x"
\end{isabelle}

With this concept at our disposal, we resume working within the locale \isa{dfa}, which is concerned with the automaton \isa{M}\@.
If no two states have the same right language, then there is a bijection between the accessible states (of \isa{M}) and the equivalence classes yielded by the relation \isa{eq\_app\_right\ language}.
\begin{isabelle}
\isacommand{lemma}\ inj\_right\_lang\_imp\_eq\_app\_right\_index:\isanewline
\ \ \isakeyword{assumes}\ "inj\_on\ right\_lang\ (dfa.states\ M)"\isanewline
\ \ \ \ \isakeyword{shows}\ "bij\_betw\ (\isasymlambda q.\ eq\_app\_right\ language\ ``\ \isacharbraceleft path\_to\ q\isacharbraceright )\isanewline
\ \ \ \ \ \ \ \ \ \ \ \ \ \ \ \ \ \ \ \ accessible\ \ (UNIV\ //\ eq\_app\_right\ language)"
\end{isabelle}
This bijection maps the state~\isa{q} to
\isa{eq\_app\_right\ language\ ``\ \isacharbraceleft path\_to\ q\isacharbraceright}.
Every element of the quotient \isa{UNIV\ //\ eq\_app\_right\ language} can be expressed in this form.
And therefore, the number of states in a \isa{minimal} machine equals the index of \isa{eq\_app\_right\ language}.
\begin{isabelle}
\isacommand{definition}\ min\_states\ \isakeyword{where}\isanewline
\ \ "min\_states\ \isasymequiv \ card\ (UNIV\ //\ eq\_app\_right\ language)"
\vskip1mm
\isacommand{lemma}\ minimal\_imp\_index\_eq\_app\_right:\isanewline
\ \ "minimal\ \isasymLongrightarrow \ card(dfa.states\ M)\ =\ min\_states"
\end{isabelle}

In the proof of the Myhill-Nerode theorem, it emerged that this index was the minimum cardinality for any DFA accepting the given language.
Any other automaton, \isa{M'}, accepting the same language cannot have fewer states.
This theorem justifies the claim that \isa{minimal} indeed characterises a minimal DFA\@.
\begin{isabelle}
\isacommand{theorem}\ minimal\_imp\_card\_states\_le:\isanewline
\ \ \ \ \ "\isasymlbrakk minimal;\ dfa\ M';\ dfa.language\ M'\ =\ language\isasymrbrakk \isanewline
\ \ \ \ \ \ \isasymLongrightarrow \ card\ (dfa.states\ M)\ \isasymle \ card\ (dfa.states\ M')"
\end{isabelle}
Note that while the locale \isa{dfa} gives us implicit access to one DFA, namely \isa{M}, it is still possible to refer to other automata, as we see above.

The minimal machine is unique up to isomorphism because every minimal machine is isomorphic to the canonical Myhill-Nerode DFA\@.
The construction of a DFA from a Myhill-Nerode relation was packaged as a locale, and by applying this locale to the given \isa{language} and the relation \isa{eq\_app\_right\ language}, we can generate the instance we need.
\begin{isabelle}
\isacommand{interpretation}\ Canon:\isanewline
\ \ \ \ MyhillNerode\_dfa\ language\ "eq\_app\_right\ language"\isanewline
\ \ \ \ \ \ \ \ \ \ \ \ \ \ \ \ \ \ \ \ \ language\ min\_states\ index\_f
\end{isabelle}
Here, \isa{index\_f} denotes some bijection between the equivalence classes and their cardinality (as an HF ordinal).
It exists (definition omitted) by the definition of cardinality itself.
It is the required isomorphism function between \isa{M} and the canonical DFA of Sect.\ts\ref{sec:canon_dfa}, which is written \isa{Canon.DFA}.
\begin{isabelle}
\isacommand{definition}\ iso\ ::\ "hf\ \isasymRightarrow \ hf"\ \isakeyword{where}\isanewline
\ \ "iso\ \isasymequiv \ index\_f\ o\ (\isasymlambda q.\ eq\_app\_right\ language\ ``\ \isacharbraceleft path\_to\ q\isacharbraceright )"
\end{isabelle}
The isomorphism property is stated using locale \isa{dfa\_isomorphism}.
\begin{isabelle}
\isacommand{theorem}\ minimal\_imp\_isomorphic\_to\_canonical:\isanewline
\ \ \isakeyword{assumes}\ minimal\ \ \ \isakeyword{shows}\ "dfa\_isomorphism\ M\ Canon.DFA\ iso"
\end{isabelle}
Verifying the isomorphism conditions requires delicate reasoning.
Hopcroft and Ullman's proof \cite[p.\ts29--30]{Hopcroft-formal} provides just a few clues.

\subsection{Brzozowski's Minimisation Algorithm}

At the core of this minimisation algorithm is an NFA obtained by reversing all the transitions of a given DFA, and exchanging the initial and final states. 
\begin{isabelle}
\isacommand{definition}\ Reverse\_nfa\ ::\ "'a\ dfa\ \isasymRightarrow \ 'a\ nfa"\ \isakeyword{where}\isanewline
\ \ "Reverse\_nfa\ MS\ =\ \isasymlparr nfa.states\ =\ dfa.states\ MS,\isanewline
\ \ \ \ \ \ \ \ \ \ \ \ \ \ \ \ \ \ \ \ \ init\ \ =\ dfa.final\ MS,\ \isanewline
\ \ \ \ \ \ \ \ \ \ \ \ \ \ \ \ \ \ \ \ \ final\ =\ \isacharbraceleft dfa.init\ MS\isacharbraceright ,\isanewline
\ \ \ \ \ \ \ \ \ \ \ \ \ \ \ \ \ \ \ \ \ nxt\ =\ \isasymlambda q\ x.\ \isacharbraceleft p\ \isasymin \ dfa.states\ MS.\ q\ =\ dfa.nxt\ MS\ p\ x\isacharbraceright ,\isanewline
\ \ \ \ \ \ \ \ \ \ \ \ \ \ \ \ \ \ \ \ \ eps\ =\ \isacharbraceleft \isacharbraceright \isasymrparr "
\end{isabelle}
This is easily shown to be an NFA that accepts the reverse of every word accepted by the original DFA\@.
Applying the powerset construction yields a new DFA that has no indistinguishable states.
The point is that the right language of a powerset state is derived from the right languages of the constituent states of the reversal NFA \cite{champarnaud-split}.
Those, in turn, are the left languages of the original DFA, and these are disjoint 
(since the original DFA has no inaccessible states, by assumption).
\begin{isabelle}
\isacommand{lemma}\ inj\_on\_right\_lang\_PR:\isanewline
\ \ \isakeyword{assumes}\ "dfa.states\ M\ =\ accessible"\isanewline
\ \ \ \ \isakeyword{shows}\ "inj\_on\ (dfa.right\_lang\ (nfa.Power\_dfa\ (Reverse\_nfa\ M)))\isanewline
\ \ \ \ \ \ \ \ \ \ \ \ \ \ \ \ \ \ (dfa.states\ (nfa.Power\_dfa\ (Reverse\_nfa\ M)))"
\end{isabelle}
The following definitions abbreviate the steps of Brzozowski's algorithm.
\begin{isabelle}
\isacommand{abbreviation}\ APR\ ::\ "'x\ dfa\ \isasymRightarrow \ 'x\ dfa"\ \isakeyword{where}\isanewline
\ \ "APR\ X\ \isasymequiv \ dfa.Accessible\_dfa\ (nfa.Power\_dfa\ (Reverse\_nfa\ X))"
\vskip1mm
\isacommand{definition}\ Brzozowski\ ::\ "'a\ dfa"\ \isakeyword{where}\isanewline
\ \ "Brzozowski\ \isasymequiv \ APR\ (APR\ M)"
\end{isabelle}

By the lemma proved just above, the \isa{APR} operation yields minimal DFAs.
\begin{isabelle}
\isacommand{theorem}\ minimal\_APR:\isanewline
\ \ \isakeyword{assumes}\ "dfa.states\ M\ =\ accessible"\isanewline
\ \ \ \ \isakeyword{shows}\ "dfa.minimal\ (APR\ M)"
\end{isabelle}
Brzozowski's minimisation algorithm is correct.
The first APR call reverses the language and eliminates inaccessible states; the second call yields a minimal machine for the original language. The proof uses the theorems just proved.
\begin{isabelle}
\isacommand{theorem}\ minimal\_Brzozowski:\ "dfa.minimal\ Brzozowski"\isanewline
\isacommand{unfolding}\ Brzozowski\_def\isanewline
\isacommand{proof}\ (rule\ dfa.minimal\_APR)\isanewline
\ \ \isacommand{show}\ "dfa\ (APR\ M)"\isanewline
\ \ \ \ \isacommand{by}\ (simp\ add:\ dfa.dfa\_Accessible\ nfa.dfa\_Power\ nfa\_Reverse\_nfa)\isanewline
\isacommand{next}\isanewline
\ \ \isacommand{show}\ "dfa.states\ (APR\ M)\ =\ dfa.accessible\ (APR\ M)"\ \isanewline
\ \ \ \ \isacommand{by}\ (simp\ add:\ dfa.Accessible\_accessible\ dfa.states\_Accessible\_dfa\isanewline
\ \ \ \ \ \ \ \ \ \ \ \ nfa.dfa\_Power\ nfa\_Reverse\_nfa)\isanewline
\isacommand{qed}
\end{isabelle}

\section{Related Work}
\label{sec:related}

There is a great body of prior work. One approach involves working constructively, in some sort of type theory.
Constable's group has formalised automata \cite{constable-constructively} in Nuprl,
including the Myhill-Nerode theorem.
Using type theory in the form of Coq and its Ssreflect library,
Doczkal et al. \cite{Doczkal-constructive} formalise much of the same material as the present paper.
They omit $\epsilon$-transitions and Brzozowski's algorithm and add the pumping lemma and Kleene's algorithm for translating a DFA to a regular expression.
Their development is of a similar length, under 1400 lines, and they allow the states of a finite automaton to be given by any finite type.
In a substantial development, Braibant and Pous \cite{braibant-deciding-kleene} have implemented a tactic for solving equations in Kleene algebras by implementing efficient finite automata algorithms in Coq.
They represent states by integers.

An early example of regular expression theory formalised using higher-order logic (Isabelle/HOL) is Nipkow's verified lexical analyser \cite{nipkow-lexical}.
His automata are polymorphic in the types of state and symbols. NFAs are included, with $\epsilon$-transitions simulated by an alphabet extended with a dummy symbol.

Recent Isabelle developments explicitly bypass automata theory.
Wu et al. \cite{Wu-formalisation} prove the Myhill-Nerode theorem using regular expressions.
This is a significant feat, especially considering that the theorem's underlying intuitions come from automata.
Current work on regular expression equivalence \cite{krauss-proof,nipkow-unified} continues to focus on regular expressions rather than finite automata.

This paper describes not a project undertaken by a team, but a six-week case study by one person.
Its successful outcome obviously reflects Isabelle's powerful automation, but the key factor is the simplicity of the specifications.
Finite automata cause complications in the prior work.
The HF sets streamline the specifications and allow elementary set-theoretic reasoning.

\section{Conclusions}
\label{sec:concl}

The theory of finite automata can be developed straightforwardly using higher-order logic and HF set theory.
We can formalise the textbook proofs: there is no need to shun automata or use constructive type theories.
HF set theory can be seen as an abstract universe of computable objects, with many potential applications.
One possibility is programming language semantics: using \isa{hf} as the type of values offers open-ended possibilities, including integer, rational and floating point numbers, ASCII characters, and data structures.

\paragraph*{Acknowledgements.}
Christian Urban and Tobias Nipkow offered advice, and suggested Brzozowski's minimisation algorithm as an example.
The referees made a variety of useful comments.

\bibliographystyle{abbrv}
\bibliography{automata}

\begin{thebibliography}{10}

\bibitem{ballarin-locales-module}
C.~Ballarin.
\newblock Locales: A module system for mathematical theories.
\newblock {\em Journal of Automated Reasoning}, 52(2):123--153, 2014.

\bibitem{braibant-deciding-kleene}
T.~Braibant and D.~Pous.
\newblock Deciding {Kleene} algebras in {Coq}.
\newblock {\em Logical Methods in Computer Science}, 8(1), 2012.

\bibitem{champarnaud-split}
J.~Champarnaud, A.~Khorsi, and T.~Parantho{\"{e}}n.
\newblock Split and join for minimizing: Brzozowski's algorithm.
\newblock In M.~Bal{\'{\i}}k and M.~Sim{\'{a}}nek, editors, {\em The Prague
  Stringology Conference}, pages 96--104. Department of Computer Science and
  Engineering, Czech Technical University, 2002.

\bibitem{constable-constructively}
R.~L. Constable, P.~B. Jackson, P.~Naumov, and J.~C. Uribe.
\newblock Constructively formalizing automata theory.
\newblock In G.~D. Plotkin, C.~Stirling, and M.~Tofte, editors, {\em Proof,
  Language, and Interaction}, pages 213--238. MIT Press, 2000.

\bibitem{Doczkal-constructive}
C.~Doczkal, J.-O. Kaiser, and G.~Smolka.
\newblock A constructive theory of regular languages in {Coq}.
\newblock In G.~Gonthier and M.~Norrish, editors, {\em Certified Programs and
  Proofs}, LNCS 8307, pages 82--97. Springer, 2013.

\bibitem{Hopcroft-formal}
J.~E. Hopcroft and J.~D. Ullman.
\newblock {\em Formal Languages and Their Relation to Automata}.
\newblock Addison-Wesley, 1969.

\bibitem{kozen-automata}
D.~Kozen.
\newblock {\em Automata and computability}.
\newblock Springer, New York, 1997.

\bibitem{krauss-proof}
A.~Krauss and T.~Nipkow.
\newblock Proof pearl: Regular expression equivalence and relation algebra.
\newblock {\em J. Autom. Reasoning}, 49(1):95--106, 2012.

\bibitem{nipkow-lexical}
T.~Nipkow.
\newblock Verified lexical analysis.
\newblock In J.~Grundy and M.~Newey, editors, {\em Theorem Proving in Higher
  Order Logics: {TPHOLs} '98}, LNCS 1479, pages 1--15. Springer, 1998.
\newblock Invited lecture.

\bibitem{nipkow-unified}
T.~Nipkow and D.~Traytel.
\newblock Unified decision procedures for regular expression equivalence.
\newblock In G.~Klein and R.~Gamboa, editors, {\em Interactive Theorem Proving
  --- 5th International Conference, {ITP} 2014}, LNCS 8558, pages 450--466.
  Springer, 2014.

\bibitem{paulson-equiv}
L.~C. Paulson.
\newblock Defining functions on equivalence classes.
\newblock {\em ACM Transactions on Computational Logic}, 7(4):658--675, 2006.

\bibitem{Finite_Automata_HF-AFP}
L.~C. Paulson.
\newblock Finite automata in hereditarily finite set theory.
\newblock {\em Archive of Formal Proofs}, Feb. 2015.
\newblock \url{http://afp.sf.net/entries/Finite_Automata_HF.shtml}, Formal
  proof development.

\bibitem{paulson-incompl-ar}
L.~C. Paulson.
\newblock A mechanised proof of {G\"odel's} incompleteness theorems using
  {Nominal Isabelle}.
\newblock {\em Journal of Automated Reasoning}, 2015.
\newblock In press. Available online at
  \url{http://link.springer.com/article/10.1007%2Fs10817-015-9322-8}.

\bibitem{swierczkowski-finite}
S.~{\'S}wierczkowski.
\newblock Finite sets and {G{\"o}del's} incompleteness theorems.
\newblock {\em Dis\-ser\-ta\-tiones Math\-e\-ma\-ti\-cae}, 422:1--58, 2003.
\newblock \url{http://journals.impan.gov.pl/dm/Inf/422-0-1.html}.

\bibitem{Wu-formalisation}
C.~Wu, X.~Zhang, and C.~Urban.
\newblock A formalisation of the {Myhill-Nerode} theorem based on regular
  expressions.
\newblock {\em Journal of Automated Reasoning}, 52(4):451--480, 2014.

\end{thebibliography}

\end{document}